\documentclass[british]{eptcs}
\usepackage[T1]{fontenc}
\usepackage[latin9]{inputenc}
\usepackage{amsthm}
\usepackage{amsmath}
\usepackage[numbers]{natbib}

\makeatletter
\numberwithin{equation}{section}
\numberwithin{figure}{section}
\theoremstyle{plain}
\newtheorem{thm}{\protect\theoremname}
  \theoremstyle{definition}
  \newtheorem{defn}[thm]{\protect\definitionname}
  \theoremstyle{plain}
  \newtheorem{lem}[thm]{\protect\lemmaname}
  \theoremstyle{plain}
  \newtheorem{prop}[thm]{\protect\propositionname}
  \theoremstyle{plain}
  \newtheorem{cor}[thm]{\protect\corollaryname}

\usepackage{verbatim}
\usepackage{stmaryrd}
\usepackage{hyperref}
\usepackage{amsmath}
\usepackage{amsfonts}
\usepackage[ruled,vlined,linesnumbered,commentsnumbered]{algorithm2e}

\long\def\beginpgfgraphicnamed#1#2\endpgfgraphicnamed{\includegraphics{#1}}
\usepackage{tikz}
\pgfrealjobname{site-graph-canonical-labelling} 
 \usetikzlibrary {shapes}
\usetikzlibrary {arrows}
\usetikzlibrary {snakes}
\usetikzlibrary {positioning}
\usetikzlibrary {decorations.pathreplacing,decorations.markings}
\usetikzlibrary {patterns}

\long\def\beginpgfgraphicnamed#1#2\endpgfgraphicnamed{\includegraphics{#1}}

\newcommand{\defsym}{~\overset{\Delta}{\simeq}~}

\newcommand{\setsize}[1]{\lvert #1 \rvert} 
\newcommand{\size}[1]{\lvert #1 \rvert} 
\DeclareMathOperator{\dom}{Dom} 
\DeclareMathOperator{\image}{Im}

\newcommand{\nodes}{V}
\newcommand{\nodesG}[1]{V_{{#1}}}
\newcommand{\nodesGext}[1]{V^\textbf{u}_{{#1}}}
\newcommand{\nodeUndef}{\textbf{Undef}} 
\newcommand{\node}{v}
\newcommand{\edges}{E}
\newcommand{\edgesG}[1]{E_{{#1}}}
\newcommand{\edge}{e}

\newcommand{\edgecolSet}{\Sigma}

\newcommand{\edgecolRel}{\prec}
\newcommand{\edgeRel}{\lhd}
\newcommand{\edgesRel}{\sqsubset}
\newcommand{\edgecolG}[1]{\phi_{#1}}
\newcommand{\edgecol}{\phi}

\newcommand{\iso}{\sigma}
\newcommand{\isorel}{\simeq}

\newcommand{\renaming}{\alpha}

\newcommand{\transG}[1]{\delta_{#1}}
\newcommand{\transGhat}[1]{\hat{\delta}_{#1}}
\newcommand{\traceeqG}[1]{\,\rho_{#1}\,}
\newcommand{\traceeq}{\,\rho\,}

\newcommand{\sitegraph}{S}
\newcommand{\sitegraphs}{\mathcal{SG}}
\newcommand{\colgraphs}{\mathcal{CG}}
\newcommand{\graphs}{\mathcal{CG}}

\newcommand{\graph}{G}
\newcommand{\sgraph}{S}
\newcommand{\canlabproc}{L}
\newcommand{\edgeenumfun}{\eta}
\newcommand{\edgeenumfunl}{{\eta_\text{L}}}

\newcommand{\isos}{I}
\newcommand{\auto}{a}

\newcommand{\res}[2]{ #1 \downarrow #2 }

\newcommand{\eqclass}[2]{ [#1]_#2 }

\newcommand{\multsetcomp}[1]{\{\hspace{-3pt}\mid #1 \mid\hspace{-3pt}\} }

\newcommand{\protnameset}{\Sigma_\text{p}}
\newcommand{\sitenameset}{\Sigma_\text{s}}
\newcommand{\protnamefun}{{\phi_\text{p}}}

\newcommand{\protnamerel}{{\,\preceq_\text{p}\,}}
\newcommand{\sitenamerel}{{\,\preceq_\text{s}\,}}
\newcommand{\site}{s}

\newcommand{\adj}{\text{Adj}}
\newcommand{\adjRev}{\text{Adj}^{-1}}

\newcommand{\totalorders}{\mathcal{T}}

\newcommand{\state}{a}
\newcommand{\sym}{x}
\newcommand{\syms}{w}
\newcommand{\statesG}[1]{A_{{#1}}}
\newcommand{\finalstatesG}[1]{F_{{#1}}}
\newcommand{\alphabetG}[1]{\Sigma_{{#1}}}

\title{Canonical Labelling of Site Graphs}

\author{Nicolas Oury$^1$
\institute{School of Informatics$^1$\\ Edinburgh University\\ Edinburgh, Scotland}
\and
\hspace{-8pt}Michael Pedersen$^{2,3}$ \qquad\quad Rasmus Petersen$^3$
\institute{ \hspace{-30pt} Department of Plant Sciences$^2$ \qquad\qquad Microsoft Research$^3$\\ 
\hspace{-22pt}Cambridge University \qquad\qquad\qquad\quad Cambridge, UK\\
\hspace{-137pt}Cambridge, UK
}
}

\def\titlerunning{Canonical Labelling of Site Graphs} 
\def\authorrunning{Nicolas Oury, Michael Pedersen \& Rasmus Petersen}

\providecommand{\DontPrintSemicolon}{\dontprintsemicolon}

\@ifundefined{showcaptionsetup}{}{%
 \PassOptionsToPackage{caption=false}{subfig}}
\usepackage{subfig}
\makeatother

\usepackage{babel}
  \providecommand{\corollaryname}{Corollary}
  \providecommand{\definitionname}{Definition}
  \providecommand{\lemmaname}{Lemma}
  \providecommand{\propositionname}{Proposition}
\providecommand{\theoremname}{Theorem}

\begin{document}
\maketitle
\begin{abstract}
We investigate algorithms for canonical labelling of site graphs, i.e. graphs in which edges bind vertices on sites with locally unique names. We first show that the problem of canonical labelling of site graphs reduces to the problem of canonical labelling of graphs with edge colourings. We then present two canonical labelling algorithms based on edge enumeration, and a third based on an extension of Hopcroft's partition refinement algorithm. All run in quadratic worst case time individually. However, one of the edge enumeration algorithms runs in sub-quadratic time for graphs with "many" automorphisms, and the partition refinement algorithm runs in sub-quadratic time for graphs with "few" bisimulation equivalences. This suite of algorithms was chosen based on the expectation that graphs fall in one of those two categories. If that is the case, a combined algorithm runs in sub-quadratic worst case time. Whether this expectation is reasonable remains an interesting open problem.
\end{abstract}

\section{Introduction}

Graphs are widely used for modelling in biology. This paper focuses
on graphs for modelling protein complexes: vertices correspond to
proteins, and edges correspond to bindings. Moreover, vertices are
labelled by protein names, and edges connect vertices on labelled
\emph{sites}, giving rise to a notion of site graphs. Importantly,
site labels can be assumed to be unique within vertices, i.e. a protein
can have at most one site of a given name. This uniqueness assumption
introduces a level of rigidity which can be exploited in algorithms
on site graphs. Rigidity is for example crucial for containing the
computational complexity of one algorithm for stochastic simulation
of rule-based models of biochemical signalling pathways \citep{danos2007scalable-simulation}. 

In this paper we investigate how rigidity can be exploited in the
design of efficient algorithms for canonical labelling of site graphs.
Informally, a canonical labelling procedure must satisfy that the
canonical labellings of two graphs are identical if and only if the
two graphs are isomorphic. A graph isomorphism is understood in the
usual sense of being an edge-preserving bijection, with the additional
requirement that vertex and site labellings are also preserved. Hence
two site graphs are isomorphic exactly when they represent protein
complexes belonging to the same species. The graph isomorphism problem
on general graphs is hard: it is not known to be solvable in polynomial
time, but curiously is not known to be NP-complete either even though
it is in NP \citep{read1977the-graph-iso-disease}. The canonical
labelling problem clearly reduces to the graph isomorphism problem,
but there is no clear reduction the other way. In this sense the canonical
labelling problem is harder than the graph isomorphism problem.

Efficient canonical labelling of site graphs has an application in
a second algorithm for stochastic simulation of rule-based languages
\citep{phillips2012modelling-engine}. This algorithm at frequent
intervals determines isomorphism of a site graph, representing a newly
created species, with a potentially large number of site graphs representing
all other species in the system at a given point in time. The algorithm
can hence compute a canonical labelling when a new species is first
created, and subsequently determine isomorphism quickly by checking
equality between this and existing canonical labellings. 

We assume in this paper that the number of site labels and the degree
of vertices in site graphs are bounded, i.e. they do not grow asymptotically
with the number of vertices, which indeed appears to be the case in
the biological setting. An algorithm for site graph \emph{isomorphism}
is presented in \citep{petrov2012site-graph-iso} and has a worst-case
time complexity of $O(\setsize{\nodes}^{2})$, where $\nodes$ is
the set of vertices. In this paper we exploit similar ideas, based
on site uniqueness, in the design of new algorithms for \emph{canonical
labelling, }which also have a worst-case time complexity of $O(\setsize{\nodes}^{2})$.
We furthermore characterise the graphs for which lower complexity
bounds are possible: if the bisimulation equivalence classes are ``small'',
meaning $O(1)$, or if the automorphism classes are ``large'', meaning
$O(\setsize{\nodes})$, then an $O(\setsize{\nodes}\cdot\log\setsize{\nodes})$
time complexity can be achieved in worst and average case, respectively. 

Site graphs can be encoded in a simpler, more standard notion of graphs
with edge-colourings. We formally define these graphs in Section \ref{sec:Site-Graphs-and-Coloured-Graphs},
along with the notion of canonical labelling and other preliminaries.
In Section \ref{sec:Edge-Enumeration-Algorithms} we specify two canonical
labelling algorithms based on ordered edge enumerations, and in Section
\ref{sec:A-Partition-Refinement} we show how these algorithms can
be improved using partition refinement. We conclude in Section \ref{sec:Conclusions}.

\section{Preliminaries\label{sec:Site-Graphs-and-Coloured-Graphs}}

\subsection{Notation}

We write $X\rightarrow Y$ (respectively $X\rightarrow_{\text{{bij}}}Y)$
for the set of total functions (respectively total bijective functions)
from $X$ to $Y$. We use the standard notation $\prod x\in X.\, T(x)$
for dependent products, i.e. the set of total functions which map
an $x\in X$ to some $y\in T(x)$. We write $\dom(f)$ and $\image(f)$
for the domain of definition and image of $f$, respectively, and
$\res{f}{\text{Z}}$ for the restriction of $f$ to the domain $Z$.
We view functions as sets of pairs $(x\mapsto y)$ when notationally
convenient. We use standard notation for finite multisets, and use
the brackets $\multsetcomp{\cdot}$ for multiset comprehension. We
write $\totalorders(X)$ for the set of total orders on $X$. We write
$X^{*}$ for the Kleene closure of $X$, i.e. the set of finite strings
over the symbols in $X$. Given a linearly ordered set $(X,<)$ we
assume the lexicographic extension of $<$ to pairs and lists over
$X$. Given a partially ordered set $(X,<)$ we write $\min_{<}(X)$
for the set of minimal elements of $X$ under $<$, and if the ordering
is total we identify $\min_{<}(X)$ with its unique least element.
Given an equivalence relation $\rho$ on a set $X$ and $x\in X$,
we write $\eqclass{x}{\rho}$ for the equivalence class of $x$ under
$\rho$, and we write $X'/\rho$ for the partition of $X'\subseteq X$
under $\rho$. Finally, given a list $x$, we write $x.i$ for the
$i$th element of $x$ starting from $1$.

\subsection{Site Graphs and Coloured Graphs}

\definecolor{nicered}{rgb}{0.75,0.3,0.3}
\definecolor{niceblue}{rgb}{0.3,0.5,0.75}
\definecolor{nicegreen}{rgb}{0.0,0.6,0.3}

\begin{figure}
\begin{centering}
\subfloat[\label{fig:site-graph}A site graph.]{

\beginpgfgraphicnamed{graph-a}
\begin{tikzpicture}[ >=latex', node distance = 2em and 3em]
  \tikzstyle{spacer} = [inner sep = 0, minimum size = 0]
  \tikzstyle{state} = [fill=#1!30, draw=black, text=black, shape = circle]
  \tikzstyle{site} = [fill=black, draw=black, inner sep = 0, minimum size = 3pt, shape = circle]
  \tikzstyle{every path} = [line width = 1pt]

  \node[state=niceblue] (n-1) {1};
  \node[state=nicered, pattern = north west lines, pattern color = nicered] (n-2) [right = of n-1] {2};
  \node[state=niceblue] (n-3) [right = of n-2] {3};

  \node[spacer, below = of n-2] (spacer) {};
	
  \node[site] (n-1-x) [right = 0pt of n-1] {};
  \node[site] (n-2-x) [left = 0pt of n-2] {};

  \node[site] (n-2-a) [above right = 0pt of n-2] {};
  \node[site] (n-3-b) [above left = 0pt of n-3] {};
  \node[site] (n-2-alpha) [below right = 0pt of n-2] {};
  \node[site] (n-3-beta) [below left = 0pt of n-3] {};

  \draw (n-1-x) to node[very near start, above]{$c$} node[very near end, above]{$c$} (n-2-x);
  \draw (n-2-a) to [bend left] node[very near start, above]{$a$} node[very near end, above]{$b$} (n-3-b);
  \draw (n-2-alpha) to [bend right] node[very near start, below]{$\beta$} node[very near end, below]{$\alpha$} (n-3-beta);
  
\end{tikzpicture} \endpgfgraphicnamed}$ $\quad\quad\quad\subfloat[\label{fig:site-graph-encoded}A coloured graph.]{

\beginpgfgraphicnamed{graph-b}
\begin{tikzpicture}[ >=latex', node distance = 2em and 3em]
  \tikzstyle{spacer} = [inner sep = 0, minimum size = 0]
  \tikzstyle{state} = [fill=white, draw=black, text=black, shape = circle, line width = 1pt]
  \tikzstyle{colloop} = [->, #1, loop above]
  \tikzstyle{every path} = [line width = 2pt]

  \node[state=niceblue] (n-1) {1};
  \node[state=nicered] (n-2) [right = of n-1] {2};
  \node[state=niceblue] (n-3) [right = of n-2] {3};

  \node[spacer, below = of n-2] (spacer) {};

  \draw[colloop=niceblue] (n-1) to (n-1);
  \draw[colloop=nicered, dotted] (n-2) to (n-2);
  \draw[colloop=niceblue] (n-3) to (n-3);
  
  \draw[->, orange, dash pattern = on 5pt off 1 pt, shorten <= 1pt] (n-1) to [bend left] (n-2);
  \draw[->, orange, dash pattern = on 5pt off 1 pt, shorten <= 1pt] (n-2) to [bend left] (n-1);
  \draw[->, nicegreen, snake = snake, line after snake = 7pt, shorten <= 1pt] (n-2) to (n-3);

\end{tikzpicture}
\endpgfgraphicnamed }\quad\quad\quad\subfloat[\label{fig:site-graph-encoded-colours}Colour definitions.]{

\beginpgfgraphicnamed{graph-c}
\begin{tikzpicture}[ >=latex', node distance = 1em and 0.5em]
  \tikzstyle{spacer} = [inner sep = 0, minimum size = 8pt, shape=rectangle]
  \tikzstyle{colnode} = [fill=#1!30, draw=black, inner sep = 0, minimum size = 8pt, shape=circle]

  \tikzstyle{colpath} = [line width = 2pt, draw=#1]

  \node[spacer] (blue) {};
  \node[right = of blue] (blueis) {=};
  \node[colnode = niceblue, right = of blueis] (bluelabel) {};

  \node[spacer, below = of blue] (red) {};
  \node[right = of red] (redis) {=};
  \node[colnode = nicered, pattern = north west lines, pattern color = nicered, right = of redis] (redlabel) {};

  \node[spacer, below = of red] (orange) {};
  \node[right = of orange] (orangeis) {=};
  \node[right = of orangeis] (orangelabel) {$\{(x_1,x_2)\}$};

  \node[spacer, below = of orange] (green) {};
  \node[right = of green] (greenis) {=};
  \node[right = of greenis] (greenlabel) {$\{(a,b), (\beta,\alpha)\}$};

  \draw[colpath = niceblue] (blue.south west) to (blue.north east);
  \draw[colpath = nicered, dotted] (red.south west) to (red.north east);
  \draw[colpath = orange, dash pattern = on 5pt off 1 pt] (orange.south west) to (orange.north east);
  \draw[colpath = nicegreen, snake=snake] (green.south west) to (green.north east);
\end{tikzpicture}
\endpgfgraphicnamed
}
\par\end{centering}

\caption{An example of a site graph (a), its encoding as a coloured graph (b),
and the formal definition of colours (c) following the encoding in
Appendix \ref{appendix:Site-Graphs} with the site name ordering $a\sitenamerel b\sitenamerel c\sitenamerel\alpha\sitenamerel\beta$.
The direction of e.g. the edge $(2,3)$ is determined by $a\sitenamerel b$
and $(a,b)$ being the minimum pair of connected sites between vertices
$2$ and $3$.}
\end{figure}
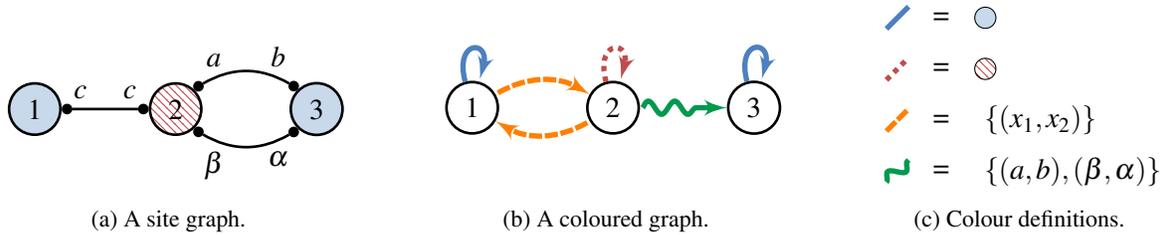

A site graph is a multi-graph with vertices labelled by protein names
and edges which connect vertices on \emph{sites }labelled by site
names. An example is shown in Figure \ref{fig:site-graph} where protein
names are indicated by colours, and a formal definition is given in
Appendix \ref{appendix:Site-Graphs}, Definition \ref{def-site-graphs}.
From an algorithmic perspective, the key property of site graphs is
their \emph{rigidity}: a vertex can have at most one site of a given
name, so a site name uniquely identifies an adjacent edge. This rigidity
property can be captured by the simpler, more standard notion of directed
graphs with edge-colourings. An example is shown in Figure \ref{fig:site-graph-encoded},
and the formal definition follows below.
\begin{defn}
Let $(\edgecolSet,\edgecolRel)$ be a given linearly ordered set of
edge colours. Then $\colgraphs$ is the set of all \emph{edge coloured
graphs} $\graph=(\nodes,\edges,\edgecol)$ where:
\begin{itemize}
\item $\nodes=\{1,\dots,k\}$ is a set of vertices.
\item $ $$\edges\subseteq\nodes\times\nodes$ is a set of directed edges.
\item $\edgecol:\edges\rightarrow\edgecolSet$ is an edge colouring satisfying
for all $\node\in\nodes$ that $\res{\edgecol}{\{(\node,\node')\in\edges\}}$
and $\res{\edgecol}{\{(\node',\node)\in\edges\}}$ are injective.
\end{itemize}
Given a coloured graph $\graph$, we write $\nodesG{\graph}$, $\edgesG{\graph}$
and $\edgecolG{\graph}$ for the vertices, edges and edge colouring
of $\graph$, respectively. We write $\adj(\node)$, respectively
$\adjRev(\node)$, for the set of all outgoing, respectively incoming,
edges adjacent to $\node$. We measure the size of graphs as the sum
of the number of edges and vertices, i.e. $\size{\graph}\defsym\size{\nodesG{\graph}}+\size{\edgesG{\graph}}$.
Since we assume the degree of vertices to be bounded, we have that
$O(\size{\graph})=O(\size{\nodesG{\graph}})=O(\size{\edgesG{\graph}})$.
\end{defn}
The condition on edge colours states that a vertex can have at most
one incident edge of a given colour, thus capturing the rigidity property.
We give an encoding of site graphs into coloured graphs in Appendix
\ref{appendix:Site-Graphs}. The encoding is injective, respects isomorphism,
and is linear in size and time (Proposition \ref{prop:The-coding-function}
in Appendix \ref{appendix:Site-Graphs}). It hence constitutes a reduction
from the site graph isomorphism problem to the coloured graph isomorphism
problem. This allows us to focus exclusively on coloured graphs in
the following. Since we are interested in protein complexes, we furthermore
assume that the underlying undirected graphs are always connected.
Figure \ref{fig:site-graph-encoded} is in fact an example encoding
of Figure \ref{fig:site-graph}, with the specific choice of colours
used by the encoding shown in Figure \ref{fig:site-graph-encoded-colours}.

\subsection{Isomorphism and Canonical Labelling}

Our notion of coloured graph isomorphism is standard. In addition
to preserving edges, isomorphisms must preserve edge colours.
\begin{defn}
Let $\graph$ and $\graph'$ be coloured graphs. An \emph{isomorphism
}is a bijective function $\iso:\nodesG{\graph}\rightarrow_{\text{{bij}}}\nodesG{\graph'}$
satisfying that $\iso(\graph)=\graph'$, i.e.:
\begin{itemize}
\item $\forall\node_{1},\node_{2}\in\nodesG{\graph}.\,[(\node,\node_{2})\in\edgesG{\graph}\Leftrightarrow(\iso(\node_{1}),\iso(\node_{2}))\in\edgesG{\graph'}]$. 
\item $\forall\edge\in\edgesG{\graph}.\,\edgecolG{\graph}(\edge)=\edgecolG{\graph'}(\iso(\edge))$.
\end{itemize}
Furthermore define $\graph\isorel\graph'$ ($\graph$ and $\graph'$
are \emph{isomorphic) }iff there exists an $\iso$ $ $relating $\graph$
and $\graph'$, and define $(\graph,\node)\isorel(\graph',\node')$
(or simply $\node\isorel\node'$) iff there exists an $\iso$ s.t.
$\iso(\node)=\node'$. We denote by $\isos(\graph,\graph')$ the set
of isomorphisms from $\graph$ to $\graph'$. An isomorphism from
$G$ to itself is called an \emph{automorphism}.
\end{defn}
Since all graphs of the same size have the same vertices, all isomorphisms
are in fact automorphisms. We next define our notion of canonical
labelling. Having vertices given by integers allows the canonical
labelling of a graph to be a graph itself.
\begin{defn}
A \emph{canonical labeller} is a function $\canlabproc:\prod\graph\in\graphs.\,\eqclass{\graph}{\isorel}$
satisfying for all $\graph,\graph'\in\graphs$ that $\canlabproc(\graph)=\canlabproc(\graph')\Leftrightarrow\graph\isorel\graph'$.
We say that $\canlabproc(\graph)$ is a \emph{canonical labelling}
of $\graph$, and that $\graph$ is \emph{canonical }if $\canlabproc(\graph)=\graph$.
\end{defn}

\section{\label{sec:Edge-Enumeration-Algorithms}Edge Enumeration Algorithms}

This section introduces two algorithms based on enumeration of edges
and the comparison of these enumerations.

\subsection{Edge Enumerators}

The rigidity property of coloured graphs, together with the linear
ordering of colours, means that any given initial node uniquely identifies
an enumeration of all the edges in a graph. Such an enumeration can
be obtained by traversing the graph from the initial node while always
following edges according to the given linear ordering on colours.
There are many possible enumeration procedures, so we generalise the
notion of an \emph{edge enumerator} as follows.
\begin{defn}
An \emph{edge enumerator }is a function of the form $\edgeenumfun:\prod\graph\in\colgraphs.\,\nodesG{\graph}\rightarrow\totalorders(\edgesG{\graph})\times(\nodesG{\graph}\rightarrow_{\text{{bij}}}\nodesG{\graph})$
satisfying for all $\graph,\graph'$ and $\node\in\nodesG{\graph},\node'\in\nodesG{\graph'}$
with $\node\isorel\node'$, $\edgeenumfun(\graph,\node)=(\edgeRel,\renaming)$
and $\edgeenumfun(\graph',\node')=(\edgeRel',\renaming')$ that $\renaming(\graph,\edgeRel)=\renaming'(\graph',\edgeRel')$.
\end{defn}
\begin{algorithm}[t]
\DontPrintSemicolon
\SetKwInOut{Input}{input}
\SetKwInOut{Output}{output}
\Input{a graph $\graph$ and a start vertex $\node\in\nodesG{\graph}$}
\Output{an enumeration of the edges in $\graph$ from $\node$}
\BlankLine
$\text{Enum} \leftarrow []$ \tcc{a list of enumerated edges}
$\renaming \leftarrow \{(\node \rightarrow 1)\}$ \tcc{a vertex renaming identifying order of encounter}
$Q \leftarrow \text{Queue}(\node)$ \tcc{a queue initialised with $\node$}
Visited $\leftarrow \{\node\}$ \tcc{a set of visited vertices}
\BlankLine
\While{$Q$ is not empty}{
	$\node \leftarrow$ Dequeue($Q$)\;
	outEdges $\leftarrow \text{Sort}_\edgecolRel(\adj(\node))$\;
	inEdges $\leftarrow \text{Sort}_\edgecolRel(\adjRev(\node))$\;
	edges $\leftarrow$ outEdges$@$inEdges\;
	\For{$i = 1$ to $\size{\emph{edges}}$}{
		$(\node_1,\node_2) \leftarrow \text{edges}.i$\;
		\If{$(\node_1,\node_2)$ does not occur in $\emph{Enum}$}{
			$\text{Enum} \leftarrow (\node_1,\node_2)::\text{Enum}$\;
			$\node_\text{new} \leftarrow \node_1 \text{ if } \node=\node_2$ and $\node_2$ if $\node=\node_1$\;
			\If{$\node_\text{new} \not\in \emph{Visited}$}{
				$\text{Visited} \leftarrow \text{Visited} \cup \{\node_\text{new}\}$\;
				$\renaming \leftarrow \renaming \cup \{(\node_\text{new} \mapsto \size{\text{Visited}})\}$\;
				Enqueue($Q,\node_\text{new}$)\;
			}
		}
	}
}
\KwRet (Enum, $\renaming$)\;
\caption{An edge enumeration algorithm. We assume given an operator $\text{Sort}_\edgecolRel(X)$ which sorts the set $X$ according to a linear ordering $\edgecolRel$. The operators $::$ and $@$ are list cons and append, respectively.}
\label{algorithm-bfs}
\end{algorithm}

An edge enumerator must hence produce a linear ordering of edges and
an alpha-conversion, with the key property that the alpha-converted
graph is invariant under isomorphism. A more direct definition is
possible, but we require the alpha-conversion to be given explicitly
by the enumerator for use in subsequent algorithms. We get the following
directly from the definition of edge enumerators.
\begin{lem}
\label{lem:edge-enum}Let $\edgeenumfun$ be an edge enumerator and
let $\edgeenumfun(\graph,\node)=(\edgeRel,\renaming)$ and $\edgeenumfun(\graph',\node')=(\edgeRel',\renaming')$
for some $\graph,\graph'$ and $\node,\node'$. If $\renaming(\graph)=\renaming(\graph')$
then $(\graph,\node)\isorel(\graph,\node')$.
\end{lem}
Algorithm \ref{algorithm-bfs} implements an edge enumerator. It essentially
carries out a breadth first search (BFS) on the underlying undirected
graph: edges are explored in order of colour, but with out-edges arbitrarily
explored before in-edges (lines $7$-$9$). The inner loop checks
whether an edge has been encountered previously (line $12$) in order
to avoid an edge being enumerated twice. The generated alpha-conversion
renames vertices by their order of discovery by the algorithm (line
$17$). 
\begin{prop}
\label{prop:edge-enumerator}Algorithm \ref{algorithm-bfs} computes
an edge enumerator.
\end{prop}
The intuition of the proof is that the control flow of the algorithm
does not depend on vertex identity, and hence the output should indeed
be invariant under automorphism. The full proof is by induction in
the number of inner loop iterations. For the complexity analysis observe
that the algorithm is essentially a BFS which hence runs in $O(\size{\nodesG{\graph}}+\size{\edgesG{\graph}})$
time, which by assumption is $O(\size{\nodesG{\graph}})$ in our case.
The extensions do not affect this complexity bound, assuming that
the checks for containment of elements in $\text{Visited}$ and $\text{Enum}$
are $O(1)$; this is possible using hash map implementations.

\subsection{A Pair-Wise Canonical Labelling Algorithm}

Alpha-converted edge enumerations can be ordered lexicographically
based on edge identity firstly, and on edge colours secondly. This
ordering is defined formally as follows.
\begin{defn}
Define a linear order $\edgesRel$ on coloured edges as $(e,c)\edgesRel(e',c')$
iff $e<e'\lor(e=e'\land c\edgecolRel c')$. We extend the order to
pairs $(\graph,e)$ s.t. $(\graph,e)\edgesRel(\graph',e')$ iff $(e,\edgecolG{\graph}(e))\edgesRel(e',\edgecolG{\graph'}(e'))$.
Finally, we assume the lexicographic extension of $\edgesRel$ to
pairs $(\graph,\edgeRel),(\graph',\edgeRel')$ of edge-ordered graphs.
\end{defn}
The ordering is used for canonical labelling by Algorithm \ref{algorithm-pairwise},
which essentially finds the edge enumeration from each vertex in the
input graph, picks the smallest after alpha-conversion, and applies
the associated alpha-conversion to the graph. Whenever two alpha-converted
enumerations are identical (line $11$), the source vertices are isomorphic
by definition of edge enumerators. The isomorphism is then computed
(line $12$) and the set of pending vertices is filtered so that it
contains at most one element from each pair of isomorphic vertices
(lines $13$-$20$). If both elements of a pair of isomorphic vertices
are in the set of pending vertices, one element is removed from the
pending set (lines $16$-$17$); the least element under $<$ is chosen
arbitrarily, which ensures that the other element does not get removed
at a later iteration. If just one element of the pair is in the pending
set (lines $18$-$19$), the other element must have been visited
earlier, and so the present pending element can be safely removed.

\begin{algorithm}[t]
\DontPrintSemicolon
\SetKwInOut{Input}{input}
\SetKwInOut{Output}{output}
\Input{a graph $\graph$ and an edge enumerator $\edgeenumfun$}
\Output{a canonical labelling of $\graph$}
\BlankLine
$\nodes_\text{pending}\leftarrow$ $\nodesG{\graph}$   \tcc{vertices yet to be enumerated }
$\edgeRel_\text{min} \leftarrow$ null \tcc{the least ordering}
$\renaming_\text{min} \leftarrow$ null \tcc{the least alpha-conversion}
\While{$\nodes_\text{pending} \ne \emptyset$}{
	$\node \leftarrow$ any $\node \in \nodes_\text{pending}$\;
	$\nodes_\text{pending} \leftarrow \nodes_\text{pending} \setminus {\node}$\;
	$(\edgeRel,\renaming) \leftarrow \edgeenumfun(G,\node)$\;
	\If{($\edgeRel_\text{min}$ = $\renaming_\text{min}$ = null) or ($\renaming(\graph,\edgeRel) \edgesRel \renaming_\text{min}(\graph, \edgeRel_\text{min})$)}{
		$\edgeRel_\text{min} \leftarrow \edgeRel$\;
		$\renaming_\text{min} \leftarrow \renaming$\;
	}
	\ElseIf{$\renaming(\graph,\edgeRel) = \renaming_\text{min}(\graph, \edgeRel_\text{min})$}{
		$\auto \leftarrow \renaming_\text{min}^{-1} \circ \renaming$ \tcc{find the automorphism }
		$\nodes'_\text{pending} \leftarrow \nodes_\text{pending}$ \tcc{keep a copy of pending vertices }
		\For{$\node' \in \dom(\renaming)$}{ \tcc{discard isomorphic vertices from pending }
			\eIf{$\{\node,\renaming(\node)\}\subseteq \nodes_\text{pending}$ }{
				$\nodes'_\text{pending} \leftarrow \nodes_\text{pending} \setminus \text{Min}_<\{\node,\alpha(\node)\}$\;
			}{
				$\nodes'_\text{pending} \leftarrow \nodes_\text{pending} \setminus \{\node,\alpha(\node)\}$\;
			}	
		}
		$\nodes_\text{pending} \leftarrow \nodes'_\text{pending}$\;
	}
}
\KwRet $\alpha_\text{min}(G)$\;
\caption{A pair-wise canonical labelling algorithm.}
\label{algorithm-pairwise}
\end{algorithm}
\begin{prop}
\label{prop:pairwise-algorithm}Algorithm \ref{algorithm-pairwise}
is a canonical labeller.
\end{prop}
The worst-case time complexity of Algorithm \ref{algorithm-pairwise}
is $O(\size{\nodesG{\graph}}^{2})$, namely when no vertices are isomorphic.
If all nodes in the input graph are isomorphic, the number of pending
vertices is halved at every iteration of the outer loop and hence
the complexity is $O(\size{\nodesG{\graph}}\cdot\log(\size{\nodesG{\graph}})).$
More generally, this bound also holds in the average case if the largest
automorphism equivalence class has size $O(\size{\nodesG{\graph}})$.
There are several ways of improving the algorithm. One way is to further
exploit automorphisms. When new automorphisms are found, these may
generate additional automorphisms through composition with existing
ones. In addition to removing elements of the pending set, automorphisms
could also be exploited by only considering the automorphism quotient
graph in subsequent iterations. Another way of improving the algorithm
is to compute edge enumerations lazily, up until the point where they
can be distinguished from the current minimum. Neither of the above
improvements, however, change the worst case complexity characteristics
of the algorithm.

\subsection{A Parallel Canonical Labelling Algorithm}

The idea of lazily enumerating edges can be taken a step further with
a second algorithm which enumerates edges from all nodes in parallel:
at each iteration, a single edge is emitted by each enumeration. Following
standard notions of lazy functions, we formally define the notion
of a \emph{lazy edge enumerator }below, where the singleton set $\{*\}$
corresponds to the \emph{unit type.}
\begin{defn}
For each coloured graph $\graph$ and $i\in\{0\dots\size{\edgesG{\graph}}-1\}$,
define the set of functions $\ensuremath{T_{i}(\graph)\defsym}$ $\{*\}\rightarrow\edgesG{\graph}\times\nodesG{\graph}\times\nodesG{\graph}\times T_{i+1}$
with $T_{\size{\edgesG{\graph}}}\defsym\colgraphs$. A \emph{lazy
edge enumerator }implementing a given enumerator $\edgeenumfun$ is
then a function $\edgeenumfunl:\prod\graph\in\colgraphs.\,\nodesG{\graph}\rightarrow T_{0}(\graph)$
satisfying for any $\graph$ and $\node\in\nodesG{\graph}$ with \mbox{$\edgeenumfun(\graph,\node)=(\edgeRel,\renaming)$}
that $(\edgesG{\graph},\edgeRel).i=e_{i}$, $\renaming(e_{i})=(\node_{i},\node_{i}')$
and $\edgeenumfunl_{\size{\edgesG{\graph}}}=\renaming(\graph)$ where
$(\edge_{i},\node_{i},\node_{i}',\edgeenumfunl_{i})\defsym\edgeenumfunl_{i-1}(*)$
for $i\in\{1\dots\size{\edgesG{\graph}}\}$ and $\edgeenumfunl_{0}\defsym\edgeenumfunl(\graph,\node)$.
\end{defn}
Hence a lazy edge enumerator produces, given a graph and a start vertex,
a function which can be applied to yield an edge, an alpha-conversion
of the edge (i.e. two vertices) and a continuation which in turn can
be applied in a similar fashion; the final function thus applied yields
an alpha-converted graph for notational convenience below. A lazy
version of the BFS edge enumerator in Algorithm \ref{algorithm-bfs}
can be implemented in a straightforward manner in a functional language.
This does not affect the complexity characteristics of the algorithm,
i.e. it still runs in $O(\size{\nodesG{\graph}})$ time.

Algorithm \ref{algorithm-parallel} computes canonical labellings
using lazy edge enumerators. It first initialises a set of enumerators
from each vertex in the input graph (line $1$). It then enters a
loop in which enumerators are gradually filtered out, terminating
when the remaining enumerators complete with an alpha-converted graph.
At termination, all remaining enumerators will have started from isomorphic
vertices, and hence they all evaluate in their last step to the same
alpha-converted graph. At each iteration, each enumerator takes a
step, yielding the next version of itself and an alpha-converted coloured
edge; the result is stored as a mapping from the former to the latter
(line $3$). A multiset of alpha-converted, coloured edges is then
constructed for the purpose of counting the number of copies of each
alpha-converted coloured edge (line $4$). The alpha-converted coloured
edges with the smallest multiplicity are selected, and of these the
least under $\edgesRel$ is selected (lines $5$-$6$). Only the enumerators
which yielded this selected edge are retained in the new set of pending
edge enumerators (line $7$). Note that further discrimination according
to connectivity with other enumerators would be possible: for example,
the vertices which are sources of enumerators eliminated in step $n$
could be distinguished from those which are sources of enumerators
eliminated in step $m\ne n$. We have omitted this for simplicity.

\begin{algorithm}
\DontPrintSemicolon
\SetKwInOut{Input}{input}
\SetKwInOut{Output}{output}
\Input{a graph $\graph$ and a lazy edge enumerator $\edgeenumfunl$}
\Output{a canonical labelling of $\graph$ }
\BlankLine
Pending $\leftarrow \{ \edgeenumfunl(\graph,\node) \mid \node \in \nodesG{\graph}\}$\;
\While{$\text{Pending} \not \subset \colgraphs$}{
	StepMap $\leftarrow \{\edgeenumfunl' \mapsto ((\node,\node'),\edgecolG{\graph}(e)) \mid (e,\node,\node',\edgeenumfunl')=\edgeenumfunl(*) \land \edgeenumfunl \in Pending\}$\;
	Steps $\leftarrow \multsetcomp{ \text{StepMap}(\edgeenumfunl') \mid \edgeenumfunl' \in \dom(\text{StepMap}) }$\;
	SmallestMult $\leftarrow \min_<\{\text{Steps}(e,c) \mid (e,c) \in \text{Steps}\}$\;
	LeastEdge $\leftarrow \min_\edgesRel\{(e,c)\in\text{Steps} \mid \text{Steps}(e,c)=\text{SmallestMult}\}$\;
	Pending $\leftarrow \{ \edgeenumfunl' \in \dom(\text{StepMap}) \mid \text{StepMap}(\edgeenumfunl')=\text{LeastEdge} \}$\;
}
\KwRet the one member of Pending\;
\caption{A parallel canonical labelling algorithm.}
\label{algorithm-parallel}
\end{algorithm}
\begin{prop}
\label{prop:parallel-algorithm}Algorithm \ref{algorithm-parallel}
is a canonical labeller.
\end{prop}
The initialisation in line $1$ runs in $O(\size{\nodesG{\graph}})$
time. The loop always requires $\size{\edgesG{\graph}}=O(\size{\nodesG{\graph}})$
iterations. In the worst case where all vertices are isomorphic, each
line within the loop requires $O(\size{\nodesG{\graph}})$ time, so
the worst case complexity is $O(\size{\nodesG{\graph}}^{2})$. Hence
in the worst case there is no asymptotic improvement over Algorithm
\ref{algorithm-pairwise}. In practice, however, Algorithm \ref{algorithm-parallel}
is likely to perform significantly better. 

The key question is how Algorithm \ref{algorithm-parallel} behaves
in cases where there are \emph{few} automorphisms, i.e. when the number
of automorphisms is sub-linear in the number of vertices. One can
hypothesise that asymmetry is then discovered sufficiently early to
yield an $O(\size{\nodesG{\graph}}\cdot\log(\size{\nodesG{\graph}}))$
time complexity. If so, an overall $O(\size{\nodesG{\graph}}\cdot\log(\size{\nodesG{\graph}}))$
algorithm is obtained by running the pairwise and the parallel algorithms
simultaneously, terminating when the first of the two algorithms terminates.
However, this question remains open. Therefore also the question of
whether a sub-quadratic time complexity bound exists in the general
case remains open.

\section{\label{sec:A-Partition-Refinement}A Partition Refinement Algorithm}

The vertex set of a coloured graph can be partitioned based on ``local
views'': vertices with the same colours of incident edges are considered
equivalent and are hence included in the same equivalence class of
the partition. If two vertices are in different classes, they are
clearly not isomorphic. The partition can then be refined iteratively:
if some vertices in a class $P$ have $c$-coloured edges to vertices
in a class $Q$ while others do not, $P$ is split into two subclasses
accordingly. This \emph{partition refinement }process can be repeated
until no classes have any remaining such diverging edges. An efficient
algorithm for partition refinement was given in 1971 by Hopcroft \citep{hopcroft1971dfa-minimisation}.
The original work was in the context of deterministic finite automata
(DFA), where partition refinement of DFA states gives rise to a notion
of language equivalence, thus facilitating minimisation of the DFA.
Hopcroft's algorithm runs in $O(n\cdot\log(n))$ where $n$ is the
number of states of the input DFA.

We show in the next subsection how our coloured graphs can be viewed
as DFA, thus enabling the application of Hopcroft's partition refinement
algorithm. The literature does provide generalisations of Hopcroft's
algorithm to other structures, including general labelled graphs \citep{Cardon82graph-partitioning}
where a vertex can have multiple incident edges with the same colour.
However, we adopt Hopcroft's DFA algorithm as this remains the simplest
for our purposes. Furthermore, the algorithm has been thoroughly described
and analysed in \citep{Knuutila2001redescribing-hopcroft}, which
we use as the basis for our presentation. In the second subsection
we extend Hopcroft's algorithm for use in canonical labelling. We
finally discuss how partition refinement relates to the edge enumeration
algorithms presented in the previous section.

\subsection{A Deterministic Finite Automata View}

We refer to standard text books such as \citep{Sipser1996ITC} for
details on automata, but recall briefly that a DFA is a tuple $(A,\Sigma,\delta,a_{0},A')$
where $A$ is a set of \emph{states}, $\Sigma$ is the \emph{input
alphabet}, $\delta:A\times\Sigma\rightarrow A$ is a total \emph{transition
function}, $a_{0}\in A$ is an \emph{initial state} and $A'\subseteq A$
is a set of \emph{final states}. A coloured graph $\graph$ is almost
a DFA: $\nodesG{\graph}$ can be taken both as the set of states and
as the set of final states, the edge colours in $\graph$ can be taken
as the input alphabet, and $\edgesG{\graph}$ determines a transition
function albeit a \emph{partial} one, meaning that the DFA is \emph{incomplete}.
There is no dedicated initial state, but initial states play no role
in partition refinement \citep{Knuutila2001redescribing-hopcroft}.
A \emph{total} transition function is traditionally obtained by adding
an additional non-accepting sink state with incoming transitions from
all states for which such transitions are not defined in the original
DFA. However, it is convenient for our purposes to add a distinct
such sink state for each state in the original DFA. This allows us
one further convenience, namely to ensure that each transition on
a colour $c$ has a reverse transition on a distinct ``reverse''
colour, $c^{-}\not\in\edgecolSet$. These ideas are formalised as
follows.
\begin{defn}
Let $\graph\in\graphs$. Define the \emph{states} $\statesG{\graph}\defsym\nodesG{\graph}\cup\nodesGext{\graph}$
where $\nodesGext{\graph}\defsym\{\nodeUndef_{\node}\mid\node\in\nodesG{\graph}\}$.
Define the \emph{final states }$\finalstatesG{\graph}=\nodesG{\graph}$.
Define the \emph{input alphabet} $\alphabetG{\graph}\defsym\image(\edgecolG{\graph})\cup\{c^{-}\mid c\in\image(\edgecolG{\graph}\}$.
Define the \emph{reversible colour transition }function $\transG{\graph}:\statesG{\graph}\times\alphabetG{\graph}\rightarrow\statesG{\graph}$
and the \emph{colour path }function\emph{ }$\transGhat{\graph}:\statesG{\graph}\times\alphabetG{\graph}^{*}\rightarrow\statesG{\graph}$
as follows:

\begin{align*}
\transG{\graph}(\state,\sym)\defsym & \begin{cases}
\node' & \,\,\text{{if}}\,\,(\state,\node')\in\edgesG{\graph}\land\edgecolG{\graph}(\state,\node')=\sym\\
\node & \,\,\text{{if}}\,\,(\node,\state)\in\edgesG{\graph}\land\edgecolG{\graph}(\node,\state)=c\land\sym=c^{-}\\
\nodeUndef_{\state} & \,\,\text{{otherwise\,\ and\,}}\state\in\nodesG{\graph}\\
\state & \,\,\text{{otherwise}}
\end{cases}\\
\transGhat{\graph}(\state,\syms)\defsym & \begin{cases}
\state & \,\,\text{{if}}\,\,\syms=\varepsilon\\
\transGhat{\graph}(\state',\syms') & \,\,\text{{if}}\,\,\syms=\sym\syms'\land\state'=\transG{\graph}(\state,\sym)
\end{cases}
\end{align*}

\end{defn}

Hence the tuple $(\statesG{\graph},\alphabetG{\graph},\transG{\graph},\finalstatesG{\graph})$
is a DFA with no initial state. It follows from rigidity of coloured
graphs and from the choice of $\nodeUndef$ states that $\transG{\graph}$
is injective. The colour path function $\transGhat{\graph}$ gives
the end state of a path specified by a colour word $w$ from an initial
state $\state$. Hence $\transGhat{\graph}(\state,\syms)\in\nodesG{\graph}=\finalstatesG{\graph}$
exactly when the word $\syms$ is accepted from the state $\state$,
or equivalently when the colour path $\syms$ exists in the graph
$\graph$. With this in mind, the following notion of \emph{vertex
bisimulation }corresponds exactly to the notion of DFA state equivalence
given in \citep{Knuutila2001redescribing-hopcroft} and proven to
be the relation computed by Hopcroft's algorithm (Corollary $15$
in \citep{Knuutila2001redescribing-hopcroft}).
\begin{defn}
Let $\graph\in\graphs$. Define the \emph{vertex bisimulation} relation
$\traceeqG{\graph}\subseteq\statesG{\graph}\times\statesG{\graph}$
as $\state\traceeqG{\graph}\state'$ iff $\forall\syms\in\alphabetG{\graph}^{*}.\,(\transGhat{\graph}(\state,\syms)\in\finalstatesG{\graph}\Leftrightarrow\transGhat{\graph}(\state',\syms)\in\finalstatesG{\graph})$.
\end{defn}

\subsection{Adapting Hopcroft's Algorithm}

The strategy of using partition refinement for canonical labelling
is to limit the number of vertices under consideration to those of
a single equivalence class in the partition $\nodesG{\graph}/\traceeqG{\graph}$.
If the selected class has size $1$, the unique vertex in this class
can be chosen as the source of canonical labelling via edge enumeration.
If the selected class is larger, one of the two edge enumeration algorithms
from Section \ref{sec:Edge-Enumeration-Algorithms} can be employed,
but starting from only the vertices of this class. 

The challenge then is how, exactly, to select an equivalence class
from the unordered partition $\statesG{\graph}/\traceeqG{\graph}$
resulting from Hopcroft's algorithm. The selection must clearly be
invariant under automorphism in order to be useful for canonical labelling.
One approach could be to employ edge enumeration algorithms on the
quotient graph $\graph/\traceeqG{\graph}$, but in the worst case
this yields quadratic time and hence defeats the purpose of partition
refinement. Instead we give an extension of Hopcroft's algorithm which
explicitly selects an appropriate class. The following definition
introduces the relevant notation, adapted from \citep{Knuutila2001redescribing-hopcroft},
required by the algorithm.
\begin{defn}
Let $\graph\in\graphs$ and let $\theta$ be an equivalence relation
on $\nodesG{\graph}$. Let $P,Q\in\statesG{\graph}/\theta$ and let
$\sym\in\alphabetG{\graph}$. Then define $P_{Q,\sym}\defsym P\cap\{\transG{\graph}^{-1}(\state,\sym)\mid\state\in Q\}$
and $P^{Q,\sym}\defsym P\setminus P_{Q,\sym}$. Also define the \emph{refiners
}$\text{{ref}}(P,\theta)\defsym\{(Q,\sym)\in(\statesG{\graph}/\theta)\times\alphabetG{\graph}\mid P_{Q,\sym}\ne\emptyset\land\size{P_{Q,\sym}}<\size{P}\}$
and the \emph{objects} $\text{{obj}}(Q,\sym,\theta)\defsym$ $\{P\in\statesG{\graph}/\theta\mid(Q,x)\in\text{ref}(P,\theta)\}$.
\end{defn}

Informally, the set $\text{{obj}}(Q,\sym,\theta)$ specifies the $\theta$-classes
which can be refined based on $\sym$-labelled transitions from states
in the class $Q$. The original version of Hopcroft's algorithm from
\citep{Knuutila2001redescribing-hopcroft} is listed in Algorithm
\ref{algorithm-hopcroft} in Appendix \ref{appendix:Site-Graphs}
for the sake of completeness. The algorithm maintains two sets: one
is the partition at a given stage (line $1$), and the second is the
set $L$ of pending refiners (line $2$), i.e. pairs of classes and
transitions which will be used for refining at a later stage. The
algorithm then loops until there are no more refiners in $L$. At
each iteration a refiner $(Q,\sym)$ is selected arbitrarily from
$L$ and used to refine all the classes in $\text{{obj}}(Q,\sym,\theta$).
The key insight of Hopcroft is to selectively add new refiners to
the set $L$, namely only the \emph{better} half of any classes which
are split and not already included as a refiner (line $14$). We choose
the smallest half to be the better one, although other choices are
possible.

The problem with the original algorithm for canonical labelling purposes
is essentially that $\statesG{\graph}/\theta$ and $L$ are maintained
as sets, hence introducing non-determinism at several points. One
solution could be to maintain both $\statesG{\graph}/\theta$ and
$L$ as lists instead, and adapt the algorithm to process their content
in a consistent order. But this would complicate the analysis and
implementation details meticulously described in \citep{Knuutila2001redescribing-hopcroft}.
Our adapted version, Algorithm \ref{algorithm-hopcroft-ext}, instead
maintains only $L$ as a list, which is not at odds with the implementation
in \citep{Knuutila2001redescribing-hopcroft}. There, $L$ is a list
of integer pairs with the first element identifying a partition and
the second element identifying an edge colour. We then change the
control flow governing $L$ to become deterministic by exploiting
the linear ordering of edge colours (lines $4$ and $13$), and otherwise
consistently operate on elements of the list (lines $7$, $15$ and
$17$). The linear ordering on edge colours is assumed extended to
states, e.g. by ordering the ``reverse'' colours after the standard
colours. Finally, we explicitly maintain a ``least'' class (line
$1$): whenever the least class is refined, we consistently choose
one sub-class to become the new least class (lines $10$ and $11$).
Note how the rigidity property is key to this extension of Hopcroft's
algorithm.

\begin{algorithm}[t]
\DontPrintSemicolon
\SetKwInOut{Input}{input}
\SetKwInOut{Output}{output}
\Input{A graph $\graph$}
\Output{The relation $\traceeqG{\graph}$ and a selected $P\in(\nodesGext{\graph}/\traceeqG{\graph})$}
\BlankLine
$M \leftarrow \nodesG{\graph}$\;
$\statesG{\graph}/\theta \leftarrow \{ \nodesG{\graph}, \nodesGext{\graph} \}$\;
$L \leftarrow []$\;
\ForEach{$\sym\in\alphabetG{\graph}$ in increasing order of $\edgecolRel$}{
	add $(\nodesGext{\graph}, \sym)$ to beginning of $L$
}
\While{$L \ne \emptyset$}{
	remove the first pair $(Q,\sym)$ from $L$\;
	\ForEach{$P\in\text{obj}(Q,\sym,\theta)$}{
		replace $P$ with $P_{Q,\sym}$ and $P^{Q,\sym}$ in $\statesG{\graph}/\theta$\;
		\If{$P=M$}{
			$M \leftarrow P_{Q,\sym}$\;
		}
	}
	\ForEach{$P$ just refined}{
		\ForEach{$\sym'\in \alphabetG{\graph}$ in increasing order of $\edgecolRel$}{
			\eIf{ $(P,\sym') \in L$}{
				replace $(P,\sym')$ with $(P_{Q,\sym},\sym')$ first and $(P^{Q,\sym},\sym')$ second in $L$\;
			}{
				AddBetter$((P_{Q,\sym},\sym')$, $(P^{Q,\sym},\sym'), \sym', L)$
			}
		}
	}
}
\KwRet $(\theta,M$)\;
\caption{An extensions of Hopcroft's partition refinement algorithm. The AddBetter routine is defined as for Algorithm \ref{algorithm-hopcroft} in Appendix \ref{appendix:Site-Graphs}.}
\label{algorithm-hopcroft-ext}
\end{algorithm}

The extensions do not affect the correctness of the algorithm because
the particular non-deterministic choices made in Hopcroft's algorithm
do not affect its final output (Corollary $10$ in \citep{Knuutila2001redescribing-hopcroft}).
The extensions do not affect the complexity analysis either, so the
algorithm still runs in $O(\size{\statesG{\graph}}\cdot\log\size{\statesG{\graph}})$
time, which is the same as $O(\size{\nodesG{\graph}}\cdot\log\size{\nodesG{\graph}})$.
In particular, the ordering of colours can be computed up front in
$O(\size{\alphabetG{\graph}}\cdot\log\size{\alphabetG{\graph}})$
by standard sorting algorithms; the list operations on $L$ can be
implemented in $O(1)$; and the comparison in line $10$ can likewise
be implemented in $O(1)$ given that the classes in $L$ can be represented
by integers.

The key property needed for canonical labelling is that the ``least''
class $M$ returned by the algorithm is invariant under automorphism.
This is indeed the case; as for the edge enumeration algorithms, the
intuition is that the control flow does not depend on vertex identity.
\begin{prop}
\label{prop:hopcroft-iso}Let $\graph,\graph'\in\graphs$ and let
$\iso\in\isos(\graph,\graph')$. Let $(\traceeq,M)$ and $(\traceeq',M')$
be the results of running Algorithm \ref{algorithm-hopcroft-ext}
on $\graph$ and $\graph'$, respectively. Then $\iso(M)=M'$.
\end{prop}
It follows that the composite algorithm which first runs partition
refinement via Algorithm \ref{algorithm-hopcroft-ext}, and then runs
one of the edge enumeration algorithms \ref{algorithm-pairwise} or
\ref{algorithm-parallel} on the returned least class, is in fact
a canonical labeller. In the cases where the selected $\traceeqG{\graph}$-class
has size $1$, the composite algorithm runs in worst case time $O(\size{\nodesG{\graph}}\cdot\log\size{\nodesG{\graph}})$.
More generally, this bound also holds if there are ``few'' bisimulations,
i.e. if $\nodesG{\graph}/\traceeqG{\graph}$ has size $O(\size{\nodesG{\graph}})$.
This is due to all bisimulation equivalence classes having the same
size as the following proposition shows, and hence the particular
choice of equivalence class does not affect time complexity.
\begin{prop}
\label{prop:partition-classes-same-size}For any coloured graph $\graph$
and any $P,Q\in(\nodesG{\graph}/\traceeqG{\graph})$, $\size{P}=\size{Q}$.
\end{prop}

\subsection{Bisimulation Versus Isomorphism}

\tikzstyle{state} = [fill, draw = black, inner sep = 0, minimum size = 5pt, shape = circle]
\tikzstyle{rededge} = [-latex, bend left, nicered, dotted, line width = 1.5pt, shorten >= 3pt, shorten <= 3pt]
\tikzstyle{blueedge} = [-latex, niceblue, solid, line width = 1.5pt, shorten <= 1pt]

\begin{figure}
\begin{centering}
\subfloat[\label{fig:graph-bisim-not-iso}]{
\beginpgfgraphicnamed{graph2-a}
\begin{tikzpicture}[node distance = 5em]
  \tikzstyle{spacer} = [inner sep = 0, minimum size = 0]

  \node[state] (n-1) {};
  \node[state] (n-2) [right = of n-1] {};
  \node[state] (n-3) [below = of n-1] {};
  \node[state] (n-4) [right = of n-3] {};

  \node[spacer, below = 3.3em of n-3] (spacer) {};

  \draw[blueedge] (n-1) to (n-2);
  \draw[blueedge] (n-2) to (n-4);
  \draw[blueedge] (n-4) to (n-3);
  \draw[blueedge] (n-3) to (n-1);
  
  \draw[rededge] (n-1) to (n-2);
  \draw[rededge] (n-2) to (n-1);
  \draw[rededge] (n-3) to (n-4);
  \draw[rededge] (n-4) to (n-3);  
\end{tikzpicture}  \endpgfgraphicnamed }$ $\quad\quad\quad\subfloat[\label{fig:big-graph}]{

\beginpgfgraphicnamed{graph2-b}
\begin{tikzpicture}[node distance = 10pt, scale = 0.53]
  \node[state] at (0,0)  (a) {};
  \node[state] at (4,-4) (b) {};
  \node[state] at (8,0)  (c) {};
  \node[state] at (4,4)  (d) {};

  \node[state] at (4,1)  (e) {};
  \node[state] at (4,-1) (f) {};
  \node[state] at (6,-1) (g) {};
  \node[state] at (6,1)  (h) {};

  \node[state] at (10,0)   (i) {};
  \node[state] at (14,4)   (j) {};
  \node[state] at (18,0)   (k) {};
  \node[state] at (14,-4)  (l) {};

  \node[state] at (14,1)  (m) {};
  \node[state] at (14,-1) (n) {};
  \node[state] at (12,-1) (o) {};
  \node[state] at (12,1)  (p) {};

  \draw[blueedge] (a) to (b); \draw[blueedge] (b) to (c);
  \draw[blueedge] (c) to (d); \draw[blueedge] (d) to (a);

  \draw[blueedge] (e) to (f); \draw[blueedge] (f) to (g);
  \draw[blueedge] (g) to (h); \draw[blueedge] (h) to (e);

  \draw[blueedge] (i) to (j); \draw[blueedge] (j) to (k);
  \draw[blueedge] (k) to (l); \draw[blueedge] (l) to (i);

  \draw[blueedge] (m) to (n); \draw[blueedge] (n) to (o);
  \draw[blueedge] (o) to (p); \draw[blueedge] (p) to (m);

  \draw[rededge] (b.north) to (f.south); \draw[rededge] (f.south) to (b.north);
  \draw[rededge] (d.south) to (e.north); \draw[rededge] (e.north) to (d.south);

  \draw[rededge] (j.south) to [bend right] (m.north); \draw[rededge] (m.north) to [bend right] (j.south);
  \draw[rededge] (l.north) to [bend right] (n.south); \draw[rededge] (n.south) to [bend right] (l.north);

  \draw[rededge] (g) to [bend left = 10] (o); \draw[rededge] (o) to [bend left = 10] (g);
  \draw[rededge] (h) to [bend left = 10] (p); \draw[rededge] (p) to [bend left = 10] (h);

  \draw[rededge] (a) to [bend left = 90] (k); \draw[rededge] (k) to [bend left = 90] (a);
  \draw[rededge] (c.east) to (i.west); \draw[rededge] (i.west) to (c.east);
\end{tikzpicture} \endpgfgraphicnamed }
\par\end{centering}

\caption{(a) A graph with one bisimulation class and two non-trivial automorphism
classes (obtained from reflection on the two diagonals). (b) A graph
with one bisimulation class and just one non-trivial automorphism
class (obtained from vertical reflection); all vertices participate
in single-coloured cycles of the same type.}
\end{figure}
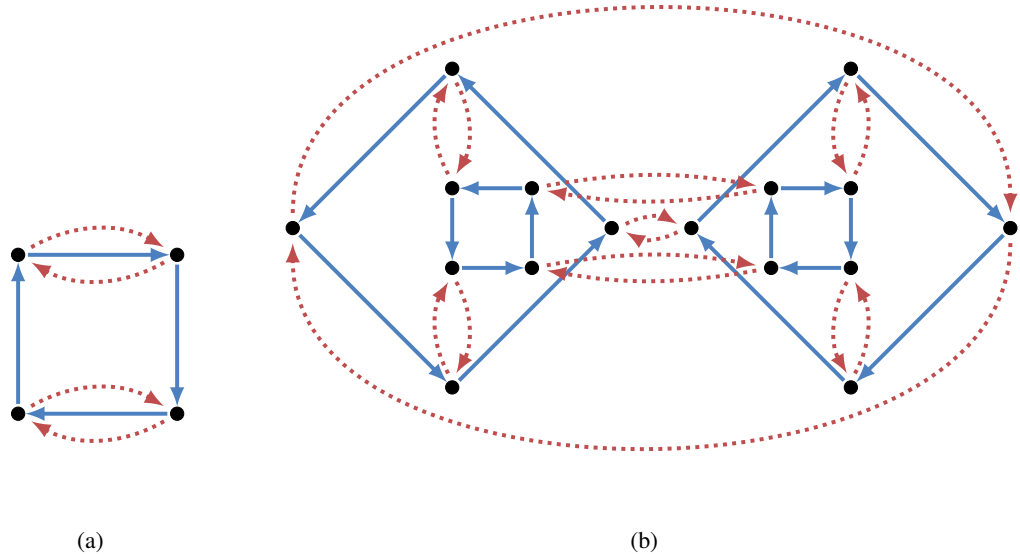

One could hope that the bisimulation and isomorphism relations for
coloured graphs were identical, for then the choice of vertex from
a selected $\traceeqG{\graph}$-class would not matter, giving a worst-case
$O(\size{\nodesG{\graph}}\cdot\log\size{\nodesG{\graph}})$ canonical
labelling algorithm. But, unsurprisingly, this is not the case. Figure
\ref{fig:graph-bisim-not-iso} shows a graph which has a single bisimulation
equivalence class but two automorphism equivalence classes. Hence
graphs of this kind, where all vertices have the same ``local view'',
capture the difficult instances of the graph isomorphism and canonical
labelling problem for site graphs. The difficulty is essentially that
isomorphisms are bijective \emph{functions }and hence must account
for vertex identity at some level. This is not the case for bisimulation
equivalence.

One possible solution could be to annotate vertices with additional,
``semi-local'' information such as edge enumeration up to a constant
length, as also suggested in \citep{petrov2012site-graph-iso}, and
then take this into account during partition refinement. However,
this is unlikely to improve the asymptotic running time. Another approach
could be to analyse the cycles of the input graph after partition
refinement, and then take this analysis into account during a second
partition refinement run on the quotient graph from the first run.
The observation here is that all same-coloured paths in the quotient
graph are cycles, and these cycles can be detected in linear time.
However, vertices with the same \emph{local} view and single-coloured
cycles are not necessarily isomorphic, as demonstrated by Figure \ref{fig:big-graph}.
It seems that simple cycles of arbitrary colour combinations must
be taken into account, e.g. to obtain cycle bases of the graph, but
there is no clear means of doing so in linear time. Further attempts
in this direction have been unsuccessful.

However, we observe that there are classes of coloured graphs for
which bisimulation and isomorphism do coincide. This holds for example
for coloured graphs with out and in-degree at most $1$, and more
generally for acyclic coloured graphs (trees). For these classes,
the partition refinement approach does yield an $O(\size{\nodesG{\graph}}\cdot\log\size{\nodesG{\graph}})$
worst case canonical labelling algorithm. Furthermore, linear time
may be possible for these classes using e.g. the partition refinement
in \citep{dovier04efficient-bisimulation}, assuming an extension
similar to that of Algorithm \ref{algorithm-hopcroft-ext} can be
realised.

Finally, we note that there is an open question of how the parallel
edge enumeration algorithm behaves on graphs with ``few'' bisimulations.
We conjecture that the algorithm may in these cases run in $O(\size{\nodesG{\graph}}\cdot\log\size{\nodesG{\graph}})$
time. If so, partition refinement would be unnecessary. However, for
the time being, the partition refinement approach does serve the purpose
of providing the upper bound of $O(\size{\nodesG{\graph}}\cdot\log\size{\nodesG{\graph}})$
for graphs with few bisimulations.

\section{Conclusions\label{sec:Conclusions}}

We have considered the problem of canonical labelling of site graphs,
which we have shown reduces to canonical labelling of standard digraphs
with edge colourings. We have presented an algorithm based on edge
enumeration which runs in $O(\size{\nodesG{\graph}}^{2})$ worst-case
time, and in $O(\size{\nodesG{\graph}}\cdot\log\size{\nodesG{\graph}})$
average-case time for graphs with many automorphisms. A variant of
this algorithm, based on parallel enumeration of edges, is likely
to perform well in praxis, but in general does not improve on the
worst case complexity bounds. However, the question of how the parallel
algorithm performs in cases with few automorphisms remains open. If
it is found to run in $O(\size{\nodesG{\graph}}\cdot\log\size{\nodesG{\graph}})$
time, this yields an overall $O(\size{\nodesG{\graph}}\cdot\log\size{\nodesG{\graph}})$
average-case algorithm and hence resolves the open question of whether
such an algorithm exists. We have also introduced an algorithm based
on partition refinement which can be used as a preprocessing step,
yielding $O(\size{\nodesG{\graph}}\cdot\log\size{\nodesG{\graph}})$
worst case time for graphs with few bisimulation equivalences. 

A different line of attack is taken in \citep{petrov2012site-graph-iso}
which introduces a notion of a graph's ``gravity centre'', namely
a subgraph on which it is sufficient to detect automorphisms. Hence
this approach is efficient when the gravity centre is small, and could
also be a useful pre-processing step for canonical labelling. However,
many of the difficult graphs that we have considered, including the
one in Figure \ref{fig:big-graph}, do not appear to have small gravity
centres.

Other related work includes that on the \emph{general} graph isomorphism
problem which has been extensively studied in the literature. Hence
highly optimised algorithms, such as the one by McKay \citep{mckay1981practical-graph-iso},
exist, although none run in sub-exponential time on ``difficult''
graphs. Many other special cases have been studied, including notably
graphs with bounded degree for which worst-case polynomial time algorithms
do exist \citep{Babai1983canonical-labelling}. Site graphs can indeed
be encoded into standard graphs with bounded degree. However, polynomial
time algorithms on such encodings do not appear to be sub-quadratic
or even quadratic. Perhaps surprisingly, isomorphism of site graphs,
or equivalently digraphs with edge colourings, has to the best of
our knowledge not been treated in the general literature. Although
the difficult cases are perhaps rare and of limited practical relevance,
the question is theoretically interesting. It certainly appears to
be conducive to infection with the graph isomorphism disease \citep{read1977the-graph-iso-disease}.

\subsection*{Acknowledgements}

We thank the anonymous reviewers for useful comments. The work of
the second author was supported by an EPSRC Postdoctoral Fellowship
(EP/H027955/1).

\bibliographystyle{plainnat}

\begin{thebibliography}{11}
\providecommand{\natexlab}[1]{#1}
\providecommand{\url}[1]{\texttt{#1}}
\expandafter\ifx\csname urlstyle\endcsname\relax
  \providecommand{\doi}[1]{doi: #1}\else
  \providecommand{\doi}{doi: \begingroup \urlstyle{rm}\Url}\fi

\bibitem[Babai and Luks(1983)]{Babai1983canonical-labelling}
L\'{a}szl\'{o} Babai and Eugene~M. Luks.
\newblock Canonical labeling of graphs.
\newblock In \emph{Proceedings of the fifteenth annual ACM symposium on Theory
  of computing}, STOC '83, pages 171--183, New York, NY, USA, 1983. ACM.
\newblock \doi{10.1145/800061.808746}.

\bibitem[Cardon and Crochemore(1982)]{Cardon82graph-partitioning}
A.~Cardon and Maxime Crochemore.
\newblock Partitioning a graph in {O}(|a| log2 |v|).
\newblock \emph{Theor. Comput. Sci.}, 19:\penalty0 85--98, 1982.
\newblock \doi{10.1016/0304-3975(82)90016-0}.

\bibitem[Danos et~al.(2007)Danos, Feret, Fontana, and
  Krivine]{danos2007scalable-simulation}
Vincent Danos, J{\'e}r{\^o}me Feret, Walter Fontana, and Jean Krivine.
\newblock Scalable simulation of cellular signaling networks.
\newblock In \emph{APLAS}, volume 4807 of \emph{LNCS}, pages 139--157.
  Springer, 2007.
\newblock \doi{10.1007/978-3-540-76637-7_10}.

\bibitem[Dovier et~al.(2004)Dovier, Piazza, and
  Policriti]{dovier04efficient-bisimulation}
Agostino Dovier, Carla Piazza, and Alberto Policriti.
\newblock An efficient algorithm for computing bisimulation equivalence.
\newblock \emph{Theor. Comput. Sci}, 311:\penalty0 221--256, 2004.
\newblock \doi{10.1016/S0304-3975(03)00361-X}.

\bibitem[Hopcroft(1971)]{hopcroft1971dfa-minimisation}
John~E. Hopcroft.
\newblock An n log n algorithm for minimizing states in a finite automaton.
\newblock Technical report, Stanford, CA, USA, 1971.

\bibitem[Knuutila(2001)]{Knuutila2001redescribing-hopcroft}
Timo Knuutila.
\newblock Re-describing an algorithm by hopcroft.
\newblock \emph{Theor. Comput. Sci.}, 250\penalty0 (1-2):\penalty0 333--363,
  January 2001.
\newblock \doi{10.1016/S0304-3975(99)00150-4}.

\bibitem[Lakin et~al.(2012)Lakin, Paulev{\'e}, and
  Phillips]{phillips2012modelling-engine}
Matthew~R. Lakin, Lo\"{\i}c Paulev{\'e}, and Andrew Phillips.
\newblock Stochastic simulation of multiple process calculi for biology.
\newblock \emph{Theor. Comput. Sci.}, 431:\penalty0 181--206, 2012.
\newblock \doi{10.1016/j.tcs.2011.12.057}.

\bibitem[McKay(1981)]{mckay1981practical-graph-iso}
B.~D. McKay.
\newblock Practical graph isomorphism.
\newblock \emph{Congr. Numer.}, 30:\penalty0 45--87, 1981.

\bibitem[Petrov et~al.(2012)Petrov, Feret, and
  Koeppl]{petrov2012site-graph-iso}
Tatjana Petrov, Jerome Feret, and Heinz Koeppl.
\newblock Reconstructing species-based dynamics from reduced stochastic
  rule-based models.
\newblock In \emph{Proceedings of the Winter Simulation Conference}, WSC '12,
  pages 1--15. Winter Simulation Conference, 2012.
\newblock \doi{10.1109/WSC.2012.6465241}.

\bibitem[Read and Cornell(1977)]{read1977the-graph-iso-disease}
R.~C. Read and D.~G. Cornell.
\newblock The graph isomorphism disease.
\newblock \emph{Journal of graph theory}, 1\penalty0 (4):\penalty0 339--363,
  1977.
\newblock \doi{10.1002/jgt.3190010410}.

\bibitem[Sipser(1996)]{Sipser1996ITC}
Michael Sipser.
\newblock \emph{Introduction to the Theory of Computation}.
\newblock International Thomson Publishing, 1st edition, 1996.
\newblock ISBN 053494728X.
\newblock \doi{10.1145/230514.571645}.

\end{thebibliography}

\appendix

\section{Site Graphs and Hopcroft's Original Algorithm \label{appendix:Site-Graphs}}

In the literature site graphs are typically defined as expressions
in a language, which is natural when considering simulation and analysis
of rule-based models. We give a more direct definition suitable for
our purposes. Site graphs in the literature often include \emph{internal
states }of sites,\emph{ }representing e.g. post-translational modification.
We omit internal states but they are straightforward to encode.
\begin{defn}
\label{def-site-graphs}Let $(\protnameset,\protnamerel)$ and $(\sitenameset,\sitenamerel$)
be given, disjoint linearly ordered sets of protein and site names,
respectively. Then $ $$\sitegraphs$ is the set of all \emph{site
graphs} $S=(\nodes,\edges,\protnamefun)$ satisfying:
\begin{itemize}
\item $\nodes=\{1,\dots,k\}$ is a set of vertices.
\item $ $$\edges\subseteq\binom{V\times\sitenameset}{2}$ is a set of site-labelled,
undirected edges satisfying that $\forall\edge,\edge'\in\edges.\,\edge\ne\edge'\rightarrow\edge\cap\edge'=\emptyset$.
\item $\protnamefun:\nodes\rightarrow\protnameset$ is a vertex (protein)
naming.
\end{itemize}
\end{defn}
Note the key condition on edges that a given site can occur at most
once within a vertex.

\begin{defn}
\label{def-site-graph-iso}Let $\sitegraph,\sitegraph'\in\sitegraphs$.
A \emph{site graph isomorphism }is a bijective function $\iso:\nodesG{\sitegraph}\rightarrow\nodesG{\sitegraph'}$
satisfying:
\begin{enumerate}
\item $\forall\node_{1},\node_{2}\in\nodesG{\sitegraph}.\,[\{(\node_{1},\site_{1}),(\node_{2},\site_{2})\}\in\edgesG{\sitegraph}\Leftrightarrow\{(\iso(\node_{1}),\site_{1}),(\iso(\node_{2}),\site_{2})\}\in\edgesG{\sitegraph'}]$. 
\item $\forall\node\in\nodesG{\sitegraph}.\,\protnamefun_{\sitegraph}(\node)=\protnamefun_{\sitegraph'}(\iso(\node))$.
\end{enumerate}
\end{defn}
The first condition states that edges and edge site names are preserved
by the isomorphism, and the second condition states that protein names
are preserved. We next show how to encode site graphs into coloured
graphs. Let $(\protnameset,\protnamerel)$ and $(\sitenameset,\sitenamerel)$
be given. The aim is to construct a linearly ordered edge colour set
$(\edgecolSet,\edgecolRel)$ and a total function $\rho:\sitegraphs(\protnameset,\sitenameset)\rightarrow\colgraphs(\edgecolSet)$
which preserves isomorphism. We define the colour set as $\edgecolSet\defsym\mathcal{{P}}(\sitenameset\times\sitenameset)\cup\protnameset$,
and the linear order on $\edgecolSet$ as $c\edgecolRel c'$ iff one
of the following conditions hold:
\begin{itemize}
\item $c\in\protnameset\land c'\not\in\protnameset$ ~or
\item $c,c'\in\protnameset\land c\protnamerel c'$ ~or
\item $c,c'\in\mathcal{{P}}(\sitenameset\times\sitenameset)\land\text{{Sort}}_{\sitenamerel}(c)\sitenamerel\text{{Sort}}_{\sitenamerel}c'$
\end{itemize}
In the latter case we assume the $\sitenamerel$ relation extended
lexicographically to pairs and lists as usual. Let $\sgraph\defsym(\nodes,\edges,\protnamefun)$
be a given site graph. We then define $\rho(\sgraph)\defsym(\nodes',\edges',\edgecol')$
where:
\begin{enumerate}
\item $\nodes'\defsym\nodes$.
\item $\edges'\defsym\edges'_{1}\cup\edges'_{2}$ where:

\begin{enumerate}
\item $\edges'_{1}\defsym\{(\node,\node')\mid\exists\site,\site'.\,\site\sitenamerel\site'\land(\site,\site')=\min_{\sitenamerel}\{(\site,\site')\mid\{(\node,\site),(\node',\site')\}\in\edges\}$
\item $\edges'_{2}\defsym\{(\node,\node)\mid\node\in\nodes\}$
\end{enumerate}
\item $\edgecol'(\node,\node')\defsym C_{1}'\cup C_{2}'$ where

\begin{enumerate}
\item $C_{1}'\defsym\{(\site,\site')\mid\{(\node,\site),(\node',\site')\}\in E\}$
\item $C_{2}'\defsym\begin{cases}
\{\protnamefun(\node)\} & \,\,\text{{if}}\,\,\node=\node'\\
\emptyset & \,\,\text{{otherwise}}
\end{cases}$
\end{enumerate}
\end{enumerate}
The encoding does not affect vertices. Note that site graphs can have
multiple unordered edges between nodes while coloured graphs have
at most one, ordered edge. The direction of this one edge is determined
in Step $2a$ from the site ordering of the least pair of sites, where
the least pair of sites is determined from the extension of the site
ordering to pairs. All vertices have self-loops (step $2b$) which
are used to encode vertex colour as edge colour in step $3b$. Step
$3$a assigns a colour to an edge as the union colours of each edge
between the edge in the site graph; the ordering of colours follows
that assigned to the edge.
\begin{prop}
\label{prop:The-coding-function}The coding function $\rho$ satisfies
the following for all $\sgraph,\sgraph'\in\dom(\rho)$:
\begin{enumerate}
\item \textbf{Injective:} $\rho(S)=\rho(S')\Rightarrow S=S'$.
\item \textbf{Respects isomorphism: }$\sgraph\isorel\sgraph'\Leftrightarrow\rho(\sgraph)\isorel\rho(\sgraph')$.
\item \textbf{Linear size: }$\size{\rho(\sgraph)}=O(\size{\sgraph})$.
\item \textbf{Linear time computable: }$\rho$ is computable in $O(\size{\sgraph})$. 
\end{enumerate}
\end{prop}
It follows immediately that the coding function together with a canonical
labeller for coloured graphs can be used to define a canonical labeller
for site graphs as follows.
\begin{cor}
Given a canonical labeller $\canlabproc:\prod\graph\in\graphs.\,[\graph]_{\isorel}$
on coloured graphs running in $O(f(\size{\graph}))\geq O(\size{\graph})$
time, the function $L^{*}(\sgraph)\defsym\rho^{-1}(\canlabproc(\rho(\sgraph)))$
is an $O(f(\size{\graph}))$ time canonical labeller on site graphs.
\end{cor}
\begin{algorithm}
\DontPrintSemicolon
\SetKwInOut{Input}{input}
\SetKwInOut{Output}{output}
\Input{a graph $\graph$}
\Output{the relation $\traceeqG{\graph}$}
$\statesG{\graph}/\theta \leftarrow \{ \nodesG{\graph}, \nodesGext{\graph} \}$\;
$L \leftarrow \emptyset$\;
\ForEach{$\sym\in\alphabetG{\graph}$}{
	add $(\nodesGext{\graph}, \sym)$ to $L$
 }
 \While{$L \ne \emptyset$}{
	remove a pair $(Q,\sym)$ from $L$\;
	\ForEach{$P\in\text{obj}(Q,\sym,\theta)$}{		replace $P$ with $P_{Q,\sym}$ and $P^{Q,\sym}$ in $\statesG{\graph}/\theta$\;
	}
	\ForEach{$P$ just refined}{
		\ForEach{$\sym'\in \alphabetG{\graph}$}{
			\If{ $(P,\sym') \in L$}{
				replace $(P,\sym')$ with $(P_{Q,\sym},\sym')$ and $(P^{Q,\sym},\sym')$ in $L$\;			
 				   }
             \Else{
				AddBetter($(P_{Q,\sym},\sym')$, $(P^{Q,\sym},\sym'), \sym', L)$
			}
		}
	}
 }
 \KwRet $\theta$\;
\BlankLine
AddBetter($P$,$Q$,$\sym$,$L$):\;
\If{$\size{P} < \size{Q}$}{
	add $(P,\sym)$ to beginning of $L$\;
 		}
 \Else{
	add $(Q,\sym)$ to beginning of $L$\;
 }
\caption{A version of Hopcroft's original algorithm adapted from Algorithm $4$ in \cite{Knuutila2001redescribing-hopcroft}.}
\label{algorithm-hopcroft}
\end{algorithm}

\end{document}